\documentclass[review, a4paper, 12pt, fleqn]{elsarticle}
\usepackage{lineno,hyperref}

\usepackage{caption}
\usepackage{subcaption}

\usepackage{multirow}

\usepackage{amsfonts}
\usepackage{amssymb}
\usepackage{amsthm}
\usepackage{verbatim}
\usepackage{graphics}
\usepackage{setspace}
\usepackage{bm}
\usepackage{lipsum}
\usepackage{xcolor}
\usepackage{ulem}

\usepackage[top = 2.7 cm, bottom = 2.7 cm, left= 2.7 cm, right = 2.7 cm]{geometry}
\nolinenumbers
\modulolinenumbers[5]

\journal{arXiv}
\begin{document}
\nolinenumbers
\begin{frontmatter}

\title{Hydrodynamic characterization of bubble column using Dynamical High Order Decomposition approach  }
%%\tnotetext[mytitlenote]{Fully documented templates are available in the elsarticle package on \href{http://www.ctan.org/tex-archive/macros/latex/contrib/elsarticle}{CTAN}.}

%%\author{Elsevier\fnref{myfootnote}}
%%\address{Radarweg 29, Amsterdam}
%%\fntext[myfootnote]{Since 1880.}

\author[cmainaddress]{Carlos Mendez}
\ead{cmendez@qui.una.py}
\author[mymainaddress]{Fabio Pereira dos Santos}
\author[gmainaddress]{Gabriel Gonçalves da Silva Ferreira}

%\ead{fsantos@eq.ufrj.br}
%\author[mymainaddress]{Ricardo de Andrade Medronho}
%\ead{medronho@eq.ufrj.br}

%%\author[mysecondaryaddress]{Global Customer Service\corref{mycorrespondingauthor}}
%%\cortext[mycorrespondingauthor]{Corresponding author}
%%\ead{support@elsevier.com}

\address[cmainaddress]{Universidad Nacional de Asunci\'{o}n, Facultad de Ciencias Qu\'{i}micas. San Lorenzo, Paraguay}
\address[mymainaddress]{Systems Engineering and Computer Science Program, Federal University of Rio de Janeiro,
Rio de Janeiro, 21941-909, Rio de Janeiro, Brazil.}
\address[gmainaddress]{Chemical Engineering Program, Federal University of Rio de Janeiro, Rio de Janeiro,
21941-972, Rio de Janeiro, Brazil.}
%%\address[mysecondaryaddress]{360 Park Avenue South, New York}

\begin{abstract}
{
Bubble columns are widely used in chemical and biochemical processes, petrochemicals, and environmental engineering industries.
Therefore, understanding bubble column dynamics is crucial to optimize their performance across various applications.
In this study, we present a data-driven approach to analyze the dynamics of a 2D bubble column system.
We conducted simulations with varying superficial velocities to create a comprehensive training dataset.
Using this data set, we compared the performance of two approaches, the Fast Fourier Transformation (FFT) and the High-Order Dynamic Mode Decomposition (HODMD), to represent and reconstruct the dynamics of the system. Our results showed that the FFT approach, which has been conventionally utilized in the industry for a long time, fails to capture the complex dispersed multiphase flow system's dynamics adequately. In contrast, HODMD successfully represented the system's dynamics with only a few sampling points.
This study highlights the potential benefits of using HODMD in analyzing bubble column dynamics, which could have significant implications for various industrial applications.}
\end{abstract}


\end{frontmatter}

%\linenumbers

\section{Introduction}

{
Bubble columns are widely used in various industries, including chemical and biochemical reactors, petrochemical, and environmental engineering. They offer numerous advantages, such as low operating costs, easy construction, high interfacial area, and efficient heat and mass transfer~\cite{KANTARCI20052263,LIU2019352,LI2022118173}. However, their complex hydrodynamics make designing and scaling-up bubble columns challenging. Accurate characterization of bubbly flows is necessary to predict the transport phenomena in multiphase flows~\cite{Shah1982353,HISSANAGA2020115531}. The gas phase is dispersed, and the liquid phase can be either continuous or in a slurry regime, which makes it difficult to simulate the behavior of bubble columns accurately.
Thus, due to their complex hydrodynamics, simulating bubble columns requires significant computational time.

Some key parameters essential for understanding, characterizing, and scaling-up of bubble columns are the gas holdup  (the fraction of gas volume in the column) and the bubble characteristics,
such as size and shape, which influence the bubble coalescence and breakup phenomena~\cite{OLMOS20016359}, as well as the flow regime~\cite{KANTARCI20052263}.
These parameters depend strongly on the superficial gas velocities, as well as other operating conditions~\cite{LIU2019352}.
Then, the resulting complex transient behavior of the bubble plume is a consequence of both the small-scale fluid-bubble and bubble-bubble interaction.
Using transient data of the velocity or gas phase fraction fields it is possible to identify the characteristic oscillation period of the flow, which is a valuable parameter to evaluate the mixing level, which affects the heat and mass transfer rates between phases~\cite{FLECK2019853}.
}

The plume dynamics has been systematically studied in the literature, either by numerical simulations or experimental results.
Most of these studies investigate the correlation between the period of the plume, the superficial velocity, the aspect ratio of the column, and the global gas hold-up~\cite{FLECK2019853,Diaz}.
In this direction,~\cite{Diaz} and~\cite{DRAHOS1991107} explored the flow regime characterization dependence with gas hold-up and the superficial gas velocity.
More specifically, \cite{Diaz} observed that the plume oscillation does not depend on the column aspect ratio for values above two.
\cite{BANNARI20083224} simulated the bubble column and characterized the plume oscillation with excellent agreement with experimental results for aspect ratio and superficial
gas velocity of $2.25$ and $0.73$ $cm/s$, respectively.

%%% Tirada de artigo ... modificar
{
\cite{DHOTRE20076615}  carried out simulations for bubble flow dynamics prediction. Their simulation gave a decent prediction of the plume oscillation, except for low-plume
oscillation regions. They conclude that a proper inter-phase momentum transfer model with the two-fluid model can obtain a satisfactory quantitative prediction of the mean flow.  
\cite{FLECK2019853} simulated several cases of oscillation bubble plume. Their simulation had a good agreement for the plume oscillation with the experimental data. They claim that the deviation is due to the lack of breakup and coalescence models. In the same path, to analyze the mixture of the phases, \cite{pr8070795} investigated the bubble oscillating dynamic process with the bubbly flow's offset behavior.
They proposed a mechanism for a no-plume oscillation period phenomenon in this sense.
}

{
Although several scientists have performed bubble column dynamics studies, there are gaps to fill in the underlying physics behind this process.
At the same time, a new wave of machine learning techniques from physics problems has gained space in the computational fluid dynamics (CFD) literature~\cite{QUEIROZ2021100002,MAIONCHI2022123110}.
However, there are only a few studies regarding the application of machine learning techniques for bubble column analysis \cite{Meisa,HAZARE2022724,Klevs}.
}

\cite{Meisa} combined CFD method and adaptive neural network to predict the flow in a bubble column reactor.
They produced a reduced-order model in which the spatial coordinates are the inputs, and the velocity field is the output.
{They obtained a good result the for velocity field, but the pressure field was not computed, which can be an essential parameter for bubble columns.}
\cite{HAZARE2022724} applied machine learning techniques to predict the gas hold-up for an industrial-scale bubble column.
The input for the training was the column diameter, the column height, the sparger design, the sparger location, the percentage free area, the superficial gas, the liquid velocity, the pressure, the temperature, the density of gas and liquid { phases, viscosity of gas and liquid and the surface tension.} They compared support vector regression, random forest, extra trees, and artificial neural network (ANN) algorithms to perform the task. They concluded that the extra tree is the best option for predicting the gas hold-up.

Another way to predict and understand the bubble column dynamics is by applying the Dynamic Mode Decomposition (DMD) \cite{Klevs,LeClainche1,Leclainche2,Leclainche3}.
\cite{Klevs} carried out DMD analysis to bubble flow with liquid-gas systems for the first time.
They applied DMD to the data set provided by a CFD simulation of a rectangular liquid metal vessel.
They obtained the flow patterns and the dynamic of the bubble plume as a proof-of-concept that the mode analysis can be used to explain the observed flow patterns.
This analysis gave them insights into momentum transfer and bubble interaction mechanisms.

In this research, we aim to provide a comprehensive understanding of the dynamics of bubble columns by utilizing Higher Order Dynamic Mode Decomposition (HODMD)\cite{LeClainche1}. Unlike traditional DMD, HODMD can extract higher-order dynamics from a dynamical system, allowing for a more in-depth analysis of the system.
Through this analysis, we not only develop a reduced-order model but also extract modal and temporal information of the system, which can be used to obtain the dominant frequency of the plume oscillation. As mentioned earlier, this is a key parameter for the bubble column design project and operation.

{
In this manuscript, we first describe the mathematical model for the bubble column simulation in Section~\ref{sec:Model}. Section~\ref{sec:DMD} summarizes the formalism for both DMD and HODMD.
The simulation results are then presented in Section~\ref{sec:Result}.
The analysis, discussion, and comparison of HODMD with Fast Fourier Transformation (FFT) are provided in Section~\ref{sec:Result}. 
Finally, we summarize our findings and their significance in Section~\ref{sec:Conclusion}.
}

\section{Mathematical modeling} \label{sec:Model}

\subsection{Governing equations}
{
The two-phase flow is modeled using the Eulerian-Eulerian approach implemented
in \texttt{twoPhaseEulerFoam} solver available in \texttt{OpenFOAM-v2012}~\cite{Weller1998}.
The mass and momentum balance equations for the liquid ($l$) and gas ($g$) phases are solved with closure models for the turbulent stress tensor and interphase forces.

The mass and momentum balance equations are the following: 
\begin{equation}
\frac{\partial \rho_k \alpha_k  }{\partial t} + \nabla \cdot \left( \rho_k \alpha_k \mathbf{u}_k  \right) = 0,
\end{equation}
\begin{equation}
    \frac{\partial \rho_k \alpha_k \mathbf{u}_k }{\partial t} + \nabla \cdot \left( \rho_k \alpha_k \mathbf{u}_k \mathbf{u}_k \right) = - \nabla \cdot \left( \alpha_k \boldsymbol{\tau}_k \right)  - \alpha_k \nabla p  + \alpha_k\rho_k\mathbf{g} + \mathbf{F}_k, \qquad k = l, g;
\end{equation}
where $\rho$ is the density, $\alpha$ is the phase fraction, $\mathbf{u}$ is the average velocity, $p$ is the average pressure field shared by all the phases, and $\mathbf{g}$ is the gravity field.
The inter-phase momentum transfer term is $\mathbf{F}_k$, and accounts for the drag, lift, virtual mass, and turbulent drag forces.
For the dispersed gas phase, the inter-phase momentum transfer term assumes the following form:
\begin{equation}
\mathbf{F}_g = \mathbf{F}_{d,g} 
	+ C_l \rho_l \mathbf{u}_r \times \left( \nabla \times \mathbf{u}_c \right)
	+ C_{vm} \rho_l \left( \frac{D_l \mathbf{u}_l}{Dt}
	- \frac{D_g \mathbf{u}_g}{Dt} \right)
	- C_{td} \alpha_g \rho_c \kappa_m \nabla \alpha_g ,
	\label{eq::interphaseForces}
\end{equation}
% Checked the drag, lift, virtual mass and dispersion terms imṕlemented within the code.
% Approach is different from described in Rusche (2003) and related papers...
% Drag approach: Symmetric formulation based on the
% described in Rusche thesis (2003) p. 103
where $\kappa_m$ is the multiphase mixture turbulent kinetic energy.
The interfacial momentum balance imposes that $\mathbf{F}_g = - \mathbf{F}_l$.
In Equation~\ref{eq::interphaseForces}, $\mathbf{F}_{d,g}$ is the drag force acting in the gas phase.
The symmetric drag formulation proposed by Weller~\cite{Weller2002} was applied:
\begin{equation}
\mathbf{F}_{d,g} = \alpha_g \alpha_l
\left( f_g \frac{C_{d,l} \rho_g}{d_g} + f_l \frac{C_{d,g} \rho_l}{d_l} \right) |\mathbf{u}_r| \mathbf{u}_r
\end{equation}
Such that:
\begin{equation}
	f_g = \min \left[ \max \left( \frac{\alpha_g - \alpha_g^{Max,f}}{\alpha_g^{Max,p} - \alpha_g^{Max,f}}, 0 \right), 1 \right] 
\end{equation}
\begin{equation}
	f_l = \min \left[ \max \left( \frac{\alpha_l^{Max,f} - \alpha_l}{\alpha_l^{Max,p} - \alpha_l^{Max,f}}, 0 \right), 1 \right] 
\end{equation}
with $\alpha_l^{Max,p} = \alpha_g^{Max,p} = 0.3$ and $\alpha_l^{Max,p} = \alpha_g^{Max,p} = 0.5$.
In this formulation $\mathbf{u}_r = \mathbf{u}_g - \mathbf{u}_l$ is the relative velocity between phases,
$C_d$, $C_l$, $C_{vm}$ and $C_{td}$ are the drag, lift, virtual mass and turbulent drag coefficients, respectively. It was considered $C_l = C_{vm} = C_{td} = 0.5$.
The Schiller and Naumann correlation~\cite{schiller1933,naumann1935} was applied for the drag coefficient:
\begin{equation}
 C_{d,g} = \max\left[ 0.44, \frac{24}{\mathrm{Re}_g} \left( 1 + 0.15 \mathrm{Re}_{g}^{0.687} \right) \right]
\end{equation}
\begin{equation}
 C_{d,l} = \max\left[ 0.44, \frac{24}{\mathrm{Re}_l} \left( 1 + 0.15 \mathrm{Re}_{l}^{0.687} \right) \right]
\end{equation}
where the Reynolds number based on the gas and liquid phase diameters, $\mathrm{Re}_g$ and $\mathrm{Re}_l$ respectively, are defined by:
\begin{equation}
 \mathrm{Re_{g}} = \frac{\rho_{l} \mathbf{u}_{r} d_g}{\mu_l} ; \qquad
 \mathrm{Re_{l}} = \frac{\rho_{g} \mathbf{u}_{r} d_l}{\mu_g}
\end{equation}
where $d$ is the dispersed phase diameter and $\mu$ is the dynamic viscosity.

The total stress tensor per unit mass is given by $\boldsymbol{\tau}_k$, and is calculated using a Reynolds-Averaged Navier-Stokes approach for the turbulence modeling of the two-phase mixture as described in \cite{behzadi2004}:
% According to the implementation of mixtureKEpsilon it appears to be the same equations
% as described in Behzadi paper.
\begin{equation}
\boldsymbol{\tau}_k = \mu_{eff,k}\left[ \nabla\mathbf{u}_k + (\nabla \mathbf{u}_k )^{T} - \frac{2}{3} \mathbf{I} \nabla\cdot\mathbf{u}_k \right],    
\end{equation}
where $\mu_{eff}$ is the effective phase viscosity, which combines the molecular and turbulent dynamic viscosity of the phase, $\mu_{eff,k} = \mu_{k} + \mu_{t,k}$.
The mixture $\kappa-\epsilon$ turbulence model equations are given as:
\begin{equation}
    \frac{\partial  \rho_m \kappa_m }{\partial t}
  + \nabla \cdot \left(  \rho_m \kappa_m \mathbf{u}_m \right) 
  - \nabla \cdot \left( \frac{\mu_{m,t}}{\sigma_m} \nabla \kappa_m \right)
  =
    G_{m, \kappa}
  - \rho_m \epsilon_m
  + S_{m, \kappa}  \ ,
\end{equation}
\begin{equation}
    \frac{\partial  \rho_m \epsilon_m }{\partial t}
  + \nabla \cdot \left(  \rho_m \epsilon_m \mathbf{u}_m \right)
  - \nabla \cdot \left( \frac{\mu_{m,t}}{\sigma_\epsilon} \nabla \epsilon_m \right)
  =
    \rho_m
		\frac{\epsilon_m}{\kappa_m}
		\left(
		C_{\epsilon,1}  G_{m, \epsilon}
	  - C_{\epsilon,2} \epsilon_m
		\right)
	  + C_{\epsilon,3} \frac{\epsilon_m}{\kappa_m} S_{m, \kappa} \ ,
\end{equation}
where $\kappa_m$ and $\epsilon_m$ are the mixture of turbulent kinetic energy and the specific turbulence dissipation rate, respectively,
$\sigma$ and $C$ are model constants, $G_{k,\kappa}$ is the turbulent kinetic energy production term and $S_{k,\kappa}$ are additional source terms that include the modeling of production
of turbulence by the presence of bubbles.
The $m$ subscript refers to the two-phase mixture properties, which are obtained from a mass-weighted average of the phase properties:
\begin{equation}
    \rho_m = \alpha_l \rho_l + \alpha_g \rho_g, \quad
    \mathbf{u}_m = \frac{\rho_l \alpha_l \mathbf{u}_l + \rho_g \alpha_g \mathbf{u}_g C_t^2}{\alpha_l \rho_l + \alpha_g \rho_g C_t^2}
\end{equation}
\begin{equation}
    \kappa_m = \frac{1}{\rho_m} \left( \alpha_l \rho_l + \alpha_g \rho_g C_t^2 \right) \kappa_l, \quad \epsilon_m = \frac{1}{\rho_m} \left( \alpha_l \rho_l + \alpha_g \rho_g  C_t^2 \right) \epsilon_l \nonumber
\end{equation}
\begin{equation}
 \mu_{m,t} = \frac{\alpha_l \mu_{l,t} + \alpha_g \mu_{g,t} C_t^2}{\alpha_l \rho_l + \alpha_g \rho_g C_t^2} \rho_m, \quad  G_{m,\kappa} = \alpha_l G_{l,\kappa} + \alpha_g G_{g,\kappa} \nonumber
\end{equation}
\begin{equation}
 S_{m,\kappa} = S_{l,\kappa} + S_{g,\kappa}
\end{equation}
More details regarding the turbulence model closure and the associated constants are found in \cite{behzadi2004}.
}

%%%%%%%%%%%%%%%%%%%%%%%%%%%%%%%%%%%%%%%%%%%%%%%%%%%%%%%%%%%%%%%%%%%%%%%%%%%%%%%%%%%%%% DMD %%%%%%%%%%%%%%%%%%%%%%%%%%%%%%%%%%%%%%%%%%%%%%%%%%%%%%
\section{Dynamical Mode Decomposition} \label{sec:DMD}

The study of the system dynamics using the DMD \cite{Schmid10} approach begins by defining the relationship between the state vector (snapshot) and the subsequent state using the linear operator ${\bf R}$ as
\begin{figure}[!htb]
\centering
\includegraphics[scale=0.6] {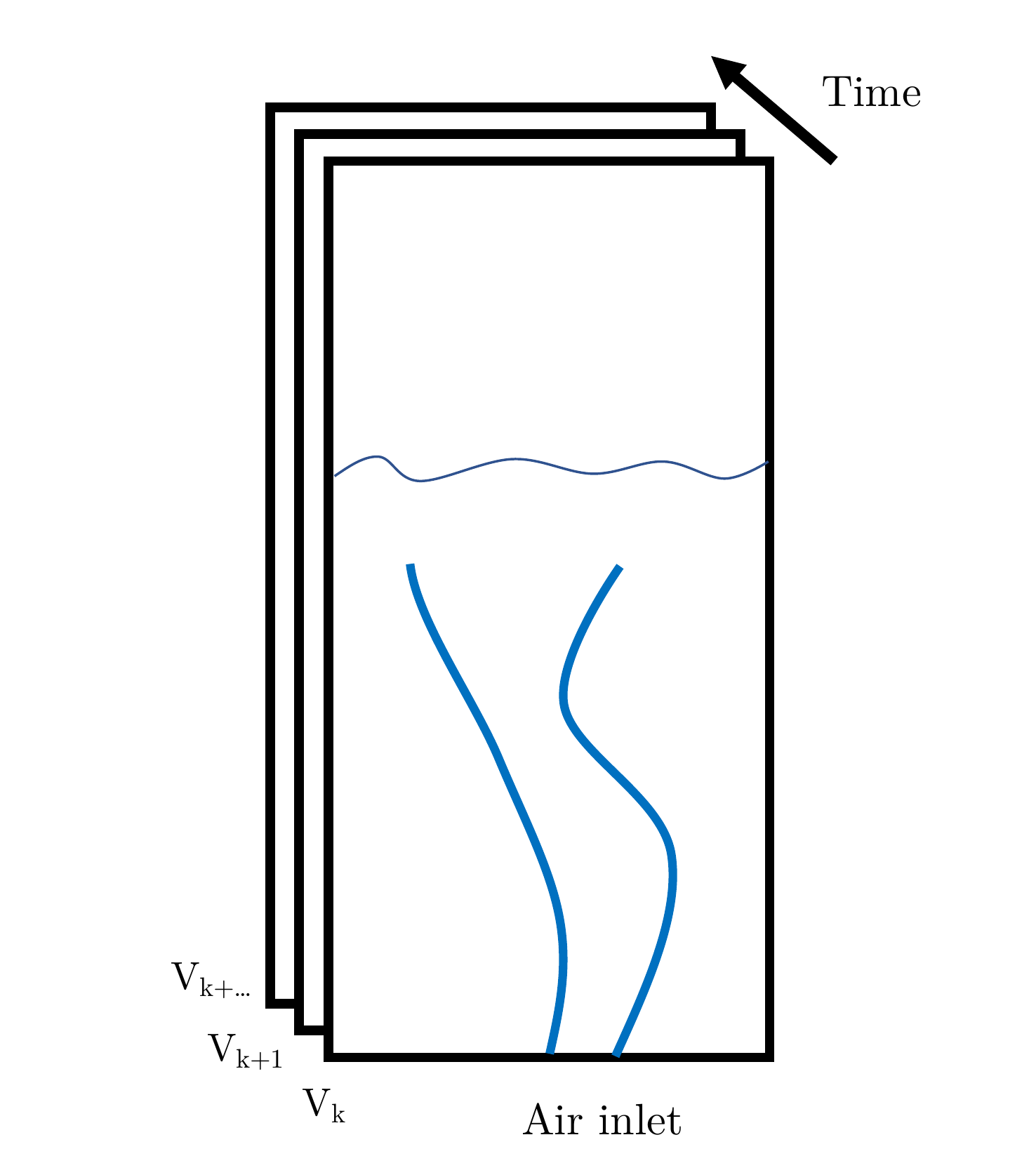}
\caption{{Data snapshots for the bubble column representation.}}
\label{fig-primeira-ordem}
\end{figure}
\begin{equation}
    \mathbf{{V}}_{k+1} \simeq \mathbf{{R}}\mathbf{{V}}_k \quad \mathrm{for } \quad k=1,\ldots,K-1,\label{b1}
\end{equation}
where the snapshots are organized in snapshots matrices of the form
\begin{equation}
    \mathbf{{V}}_p^Q = [\mathbf{v}_{p},\mathbf{v}_{p+1},\ldots,\mathbf{v}_{k},\mathbf{v}_{k+1},\ldots,\mathbf{v}_{Q-1},\mathbf{v}_{Q}]. \label{b11}
\end{equation}

The system's dynamics (damping rates and frequencies) are determined from the eigenvalues of ${\bf R}$.
The HODMD improves the results by cleaning the noise and removing the spatial redundancies, as will be explained in Subsection \ref{sec:HODMD}.
In temporal signals, it is possible to decompose periodic or quasi-periodic raw data $\mathbf{v}_k$ as an expansion of $M$ Fourier-like modes:
\begin{equation}
    \mathbf{v}_{k}\simeq  \mathbf{v}_{k}^{approx.}\equiv
    \sum_{m=1}^M a_{m} {\bf r}_m e^{(\zeta_m+i \omega_m)(k-1)\Delta t} ,\quad
    k=1,\ldots,K,\label{b0}     
\end{equation}
where the number of terms $M$ is the \textit{spectral complexity}, $K$ is the \textit{temporal dimension}, ${\bf r}_m$ are the spatial coefficients or modes (related to stationary patterns for different fields), and $a_m$ are their corresponding amplitudes.
The dimension of the subspace generated by the $M$ modes is the \textit{spatial complexity} $N$.
In the next subsection, the application of the HODMD to the simulation data is explained.

\subsection{Higher order dynamic mode decomposition}\label{sec:HODMD}
%\vspace{-4pt}
The main goal of HODMD is to express a set of instantaneous (Spatio-temporal) data as an expansion of modes as shown in Equation~\ref{b0}; thus, it is possible to study the main frequencies and growth rates composing a signal.
{ For convenience, consider two sets of $K$ time-equispaced snapshots (with $\Delta t$) obtained in the simulation, named $\mathbf{U}_1^K$ and $\mathbf{P}_1^K$, being the normalized pressure and velocity fields, respectively}
\begin{equation}
    \mathbf{U}_1^K = [\mathbf{u}_{1},\mathbf{u}_{2},\ldots,\mathbf{u}_{k},\mathbf{u}_{k+1},\ldots,\mathbf{u}_{K-1},\mathbf{u}_{K}]
\end{equation}
\begin{equation}
    \mathbf{P}_1^K = [\mathbf{p}_{1},\mathbf{p}_{2},\ldots,\mathbf{p}_{k},\mathbf{p}_{k+1},\ldots,\mathbf{p}_{K-1},\mathbf{p}_{K}]
\end{equation}

{Each vector $\bf{u_{k}}$ and $\bf{p_{k}}$ corresponds to the field values obtained by the CFD model at given time $t_k$.
As each vector corresponds to the representative values of the field for every mesh volume centroid, the dimension of this matrix is $J \times K$, where $J$ is the number of volumes of the mesh.}
The HODMD algorithm is explained in two main steps \cite{LeClainche1}
\begin{enumerate}
\item {\it Dimension reduction of the system} \\
Reduce the original data $J$ (the number of nodes of the mesh) to a set of linearly dependent vectors of dimension $N$, reducing the noise of the signal.
In this way, singular value decomposition (SVD) is applied to the snapshots matrix, obtaining  $N\times N$ unit matrix and the diagonal of matrix $\boldsymbol{\Sigma}$ contains the singular values $\sigma_1,\ldots,\sigma_{K}$.
The number of retained SVD modes, $N$, is calculated through the standard SVD error estimated for a certain tolerance $\varepsilon_1$.
\item {\it DMD-d} \\
{
The following higher-order Koopman assumption is applied to the pressure and velocity fields:
\begin{equation}
    \mathbf{{U}}^{K}_{d+1} \simeq \mathbf{{R}}_1 \mathbf{{U}}^{K-d}_1 + \mathbf{{R}}_2 \mathbf{{U}}^{K-(d-1)}_2 +...+\mathbf{{R}}_d \mathbf{{U}}^{K-1}_d \label{b4a}
\end{equation}
\begin{equation}
    \mathbf{{P}}^{K}_{d+1} \simeq \mathbf{{R}'}_1 \mathbf{{P}}^{K-d}_1 + \mathbf{{R}'}_2 \mathbf{{P}}^{K-(d-1)}_2 + ... + \mathbf{{R'}}_d \mathbf{{P}}^{K-1}_d \label{b5a}
\end{equation} 
is applied to the reduced snapshots matrix.
The eigenvalues of the \textit{modified Koopman matrix} $\mathbf{U}^{K-d}_{k+1} \simeq \mathbf{R} \mathbf{U}^{K-d+1}_k$ (which contain all the Koopman operators of the Eq~\ref{b4a}) gives the frequencies and damping rates of the DMD expansion Equation~\ref{b0}.
The eigenvectors are used to calculate the DMD modes ${\bf r}_m$.
Finally, the amplitudes related to each mode are calculated using the least squares fitting. 
The user sets a second tolerance $\varepsilon_2$ to retain the most relevant HODMD modes, as a function of their amplitudes.
The parameter $d$ identified in Equation~\ref{b4a} is calibrated for the results' robustness (similar results for different $d$ and $\Delta t$).
A more detailed description of this algorithm is presented in the literature \cite{LeClainche1}.
}

\end{enumerate}

\subsection{HODMD of the complete field}

Before presenting the results of frequency, decay rate, amplitude, and modal shape of the modes, it is necessary to explain the methodology used \cite{MendezHODMD}.
\begin{figure}[!htb]
\centering
\includegraphics[scale=0.5] {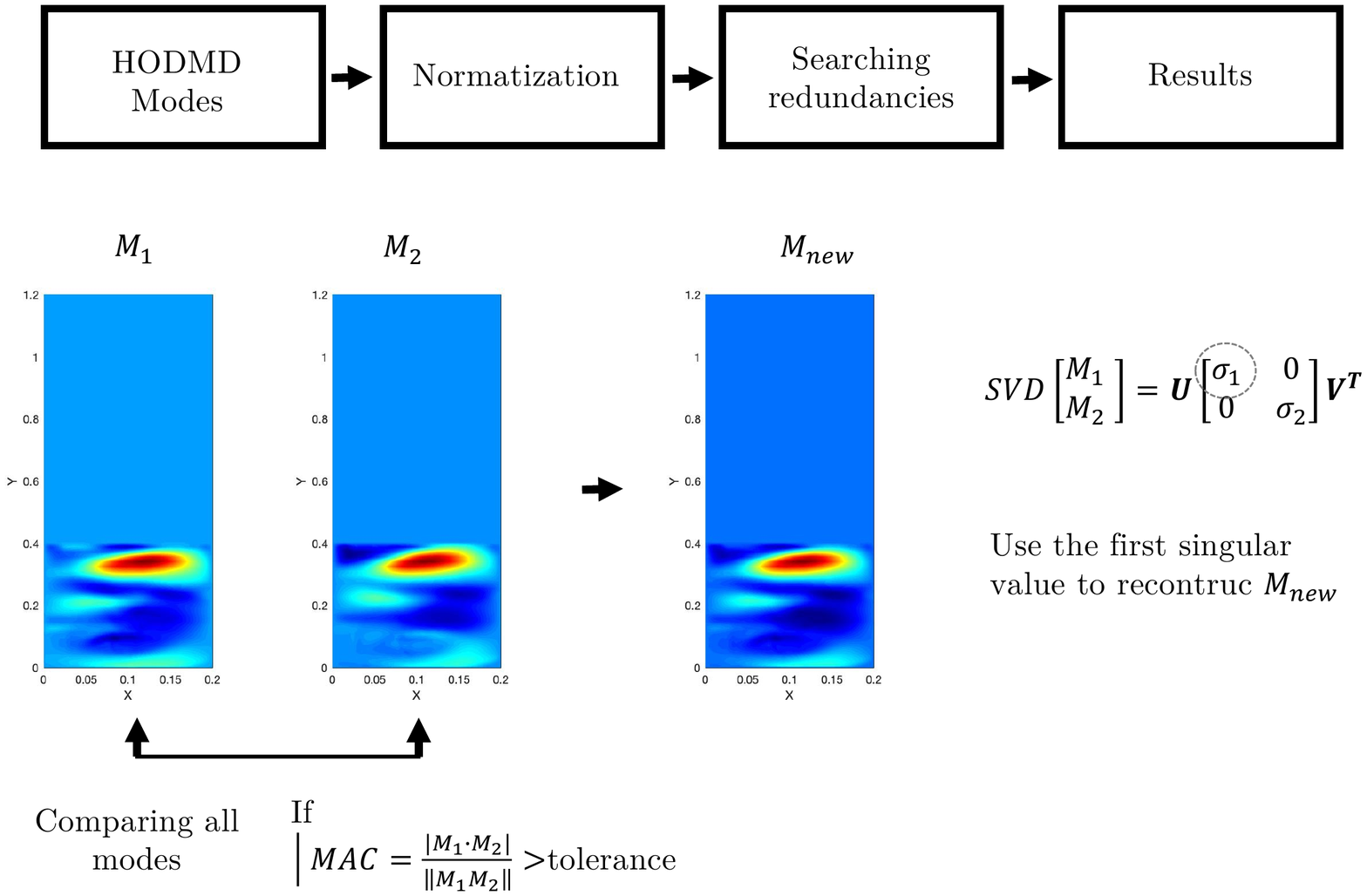}
\caption{Esquematic representation of the proposed algorithm.}
\label{fig::figHODMD_Steps}
\end{figure}

Figure~\ref{fig::figHODMD_Steps}, shows the stages carried out to obtain the results. These steps include:
\begin{itemize}
\item \textbf{HODMD algorithm application:} In this stage, a truncation value \textit{N} (related to $\varepsilon_1$), a value of \textit{d}, and the threshold $\varepsilon_2$ are selected.
The final result is a set of pairs (real and imaginary) of the most relevant modes of the whole field section).
\item  \textbf{Normalization}: This step prepares the set of modes for comparison by normalizing them relative to the maximum.
\item  \textbf{Redundancy search}: For complex problems, it is possible that, given a selection of thresholds, spurious or redundant modes are obtained (same modal shape and frequency); therefore, to avoid this, all modes are compared with each other using the Modal Assurance Criterion (MAC) explained in \cite{MendezHODMD}.
In the case of similar modes, the procedure consists of the decomposition in singular values (SVD) of the Modes $M_1$ and $M_2$ (see Figure~\ref{fig::figHODMD_Steps}).
This decomposition gives two singular values; the first is selected and used to reconstruct the full mode $M_{new}$.
\item \textbf{Results:} Finally, the obtained modes are presented in some figures, as shown in the following section.
\end{itemize}

It is important to note that our methodology is focused on the liquid phase, where its dynamical system is relevant, for which it is applied to the normalized pressure and x-velocity fields.

%\begin{table}[!htb]
%\centering
%\begin{tabular}{ r r r }
%\hline \hline
%\multicolumn{3}{c}{ Frequency - FFT} \\ \hline \hline
%  Velocity (m/s) & $Freq_p$ (Hz) &	$Freq_v$ (Hz)  \\
%  0.14 &  0.033 & 0.036   \\
%  0.50 &  0.040 & 0.040   \\
%  0.73 &  0.060 & 0.060   \\
%  0.83 &  0.14 & 0.075   \\
%\hline
%\end{tabular}
%\caption{Frequencies obtained using FFT algorithm for velocity ($U_x$) and pressure ($P$) field.}
%\label{tab:StatistcMea}
%\end{table}

\section{Simulations results}\label{sec:Result}
\begin{figure}
     \begin{subfigure}[b]{0.45\textwidth}
         \centering
         \includegraphics[scale=0.35]{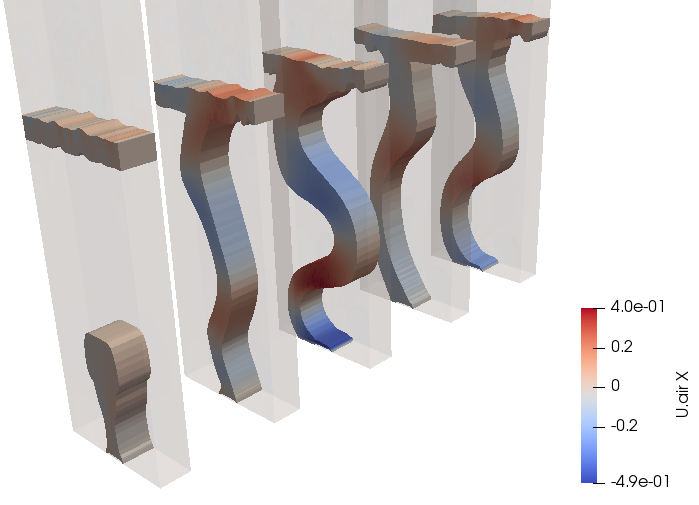}
         \caption{Isovolume of the plume dynamics over time colored by air velocity in x direct for inlet superficial velocity of $0.5~cm/s$.}
         \label{fig:isoVel}
     \end{subfigure}
     \hfill
     \begin{subfigure}[b]{0.45\textwidth}
         \includegraphics[scale=0.32]{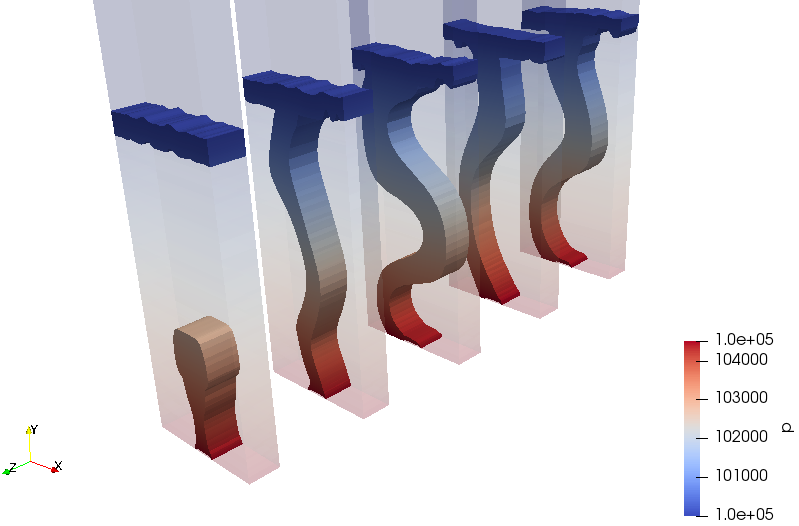}
        \caption{Isovolume of the plume dynamics over time colored by pressure for inlet superficial velocity of $0.5~cm/s$.}
         \label{fig:isoPress}
     \end{subfigure}
\end{figure}

{
The bubble column simulations were carried out using a two-dimensional model with a width of $20~cm$ and a height of $120~cm$.
 The column is filled with a water column  with an initial height of $20~cm$ ($\rho_l=1027~kg/m^3$ and $\mu_l=3.645\times 10^{-4}~Pa.s$).
Air is injected through the bottom of the column, specifically in the center position, through a $36~mm$ width inlet ($\rho_g = 1.27~kg/m^3$ and $\mu_g = 1.84\times 10^{-5}~Pa.s$).
 {It has an equivalent rectangular region of $36 \times 5~mm$ size in the simulations.}
The gas injection superficial velocity is prescribed in the inlet boundary condition, and the gas phase fraction is set to 1 at this boundary.
An outlet boundary condition is imposed on the top of geometry, which maintains a specified pressure of the internal domain.
On the walls, a no-slip condition is imposed for both fluids.
The bubble diameter is set to a fixed value of $1.3~mm$.
As mentioned above, all CFD simulations were performed using the \texttt{twoPhaseEulerFoam} solver for the numerical solution.
For spatial discretization, we applied the Van Leer convective scheme, and a Gauss linear scheme for the Laplacian-like and gradient terms.
For temporal integration, we used the implicit Euler methods with an adaptive time step to ensure that the maximum Courant-Friedrichs-Lewy (CFL) number remained below 0.1 for all cases.
}
% Informação de geometria 20 cm base e  me1.2tros altura
% 2D simulation total time of 150s steady state plume
% Hold up de liquid inicial 45 cm
% 12600 nodesu used
% 120x(2*30+20)
% 100x(2*25 + 15)  with non-uniform mesh - used
% 60x(2*15+8) coarse 6360 nodes
%% Boudary conditions
{
Figures~\ref{fig:isoVel} and~\ref{fig:isoPress} show the plume dynamics over time in each period of oscillation for a gas superficial velocity of $0.5~cm/s$.
Figure~\ref{fig:isoVel} displays an iso-volume of air volume fraction above 0.5, colored by air velocity.
It is evident from the figure that the gas phase oscillates primarily in the $x$ direction.
This observation motivates us to analyze the characteristics of the system using this variable.

Figure~\ref{fig:isoPress} illustrates the plume dynamics colored by the pressure field.
While there are also oscillations in the pressure field, these are more difficult to observe due to their relatively low magnitude variation.
Despite this, the pressure field plays a critical role in the hydrodynamics of the system and can significantly impact the behavior of the gas-liquid flow.
In the next section, we will present the results of the data-driven hydrodynamic characterization of the system.
}

\section{Dynamic Capture Results}\label{sec:Resultdmd}
{
To carry out the planned analyses for this study, we performed simulations at various air inlet velocities: 0.14, 0.23, 0.43, 0.50, 0.66, 0.73, 0.83, and 0.90 cm/s. The setup and modeling of the simulation were explained in the previous section.

\subsection{Fast Fourier Transformation (FFT) analysis}

In bubble column studies, the standard procedure for flow characterization typically involves the application of Fast Fourier Transform (FFT) analysis at a selected point within the column. However, the accuracy and reliability of this approach depend heavily on the expertise of the researcher in selecting an appropriate location that can provide representative results of the system dynamics, particularly those related to plume oscillation.}

\begin{figure}[!htb]
\centering
\includegraphics[scale=0.094] {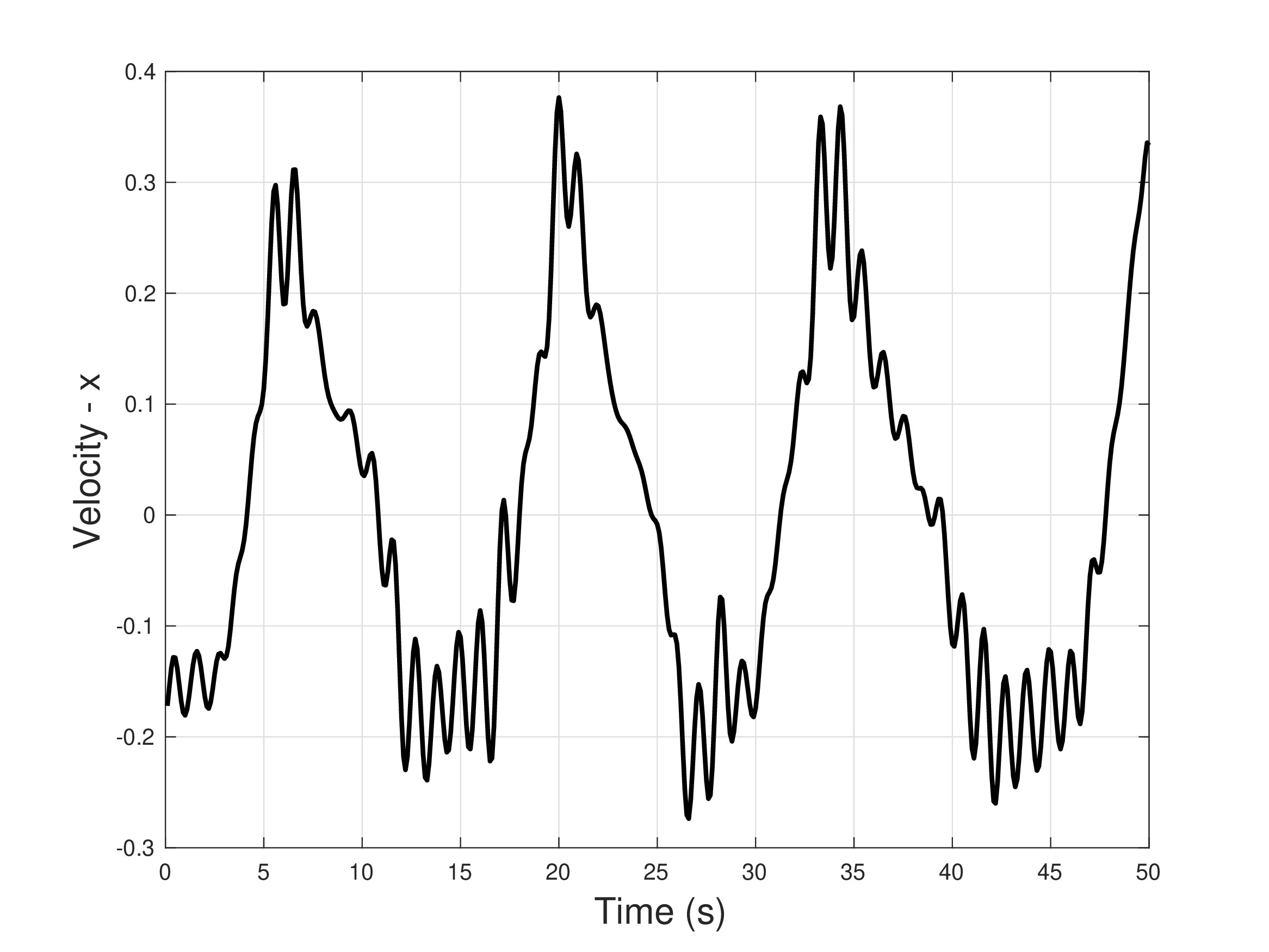}
\includegraphics[scale=0.094] {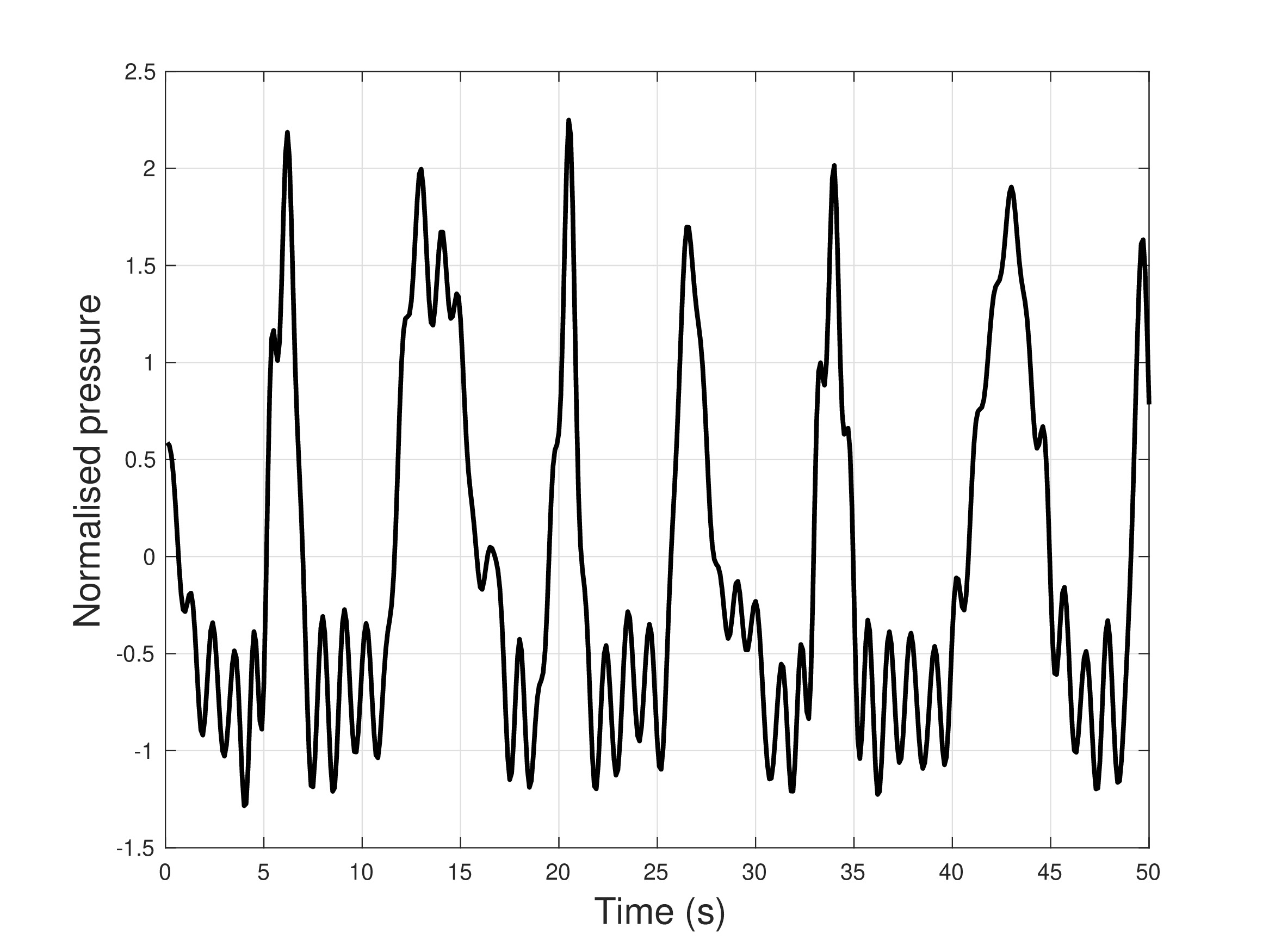}
\caption{Air velocity in X and pressure (normalized) values over time for a representative point in the middle of the liquid phase.}
\label{fig:fftResults}
\end{figure}

{
The processing applied to the signals, such as normalization, detrending, and low-pass filtering, was performed according to the procedures outlined in \cite{BUWA20024715}.
Figure~\ref{fig:fftResults} shows that all the variables present a periodic behavior, which, in essence, is favorable for applying techniques such as HODMD.
Evaluating the transient behavior of the velocity components, the magnitude of the Y direction is higher than that of the X direction, as expected for a bubble column.
However, since the primary interest of studying the hydrodynamics of bubble columns is to understand the plume's behavior, the analysis is focused on the dynamic decomposition of the X-direction velocity.
Additionally, the normalized pressure signal exhibits a favorable behavior, making it an essential variable for the proposed methods in this paper. It is noteworthy that pressure signals have practical applications since it is relatively easier to find and couple pressure sensors or transducers to the system when compared to the velocity of volumetric fraction fields.
}

{
As shown in Figure~\ref{fig::fftSpectrum} below, the FFT applied to a point in the pressure field produces a frequency spectrum with several peaks, which is characteristic of the system behavior.
However, identifying the dominant frequency is crucial for hydrodynamic characterization purposes, and its value is highly dependent on the choice of the correct analysis point for the FFT application.
}
\begin{figure}[!htb]
\centering
\includegraphics[scale=0.25] {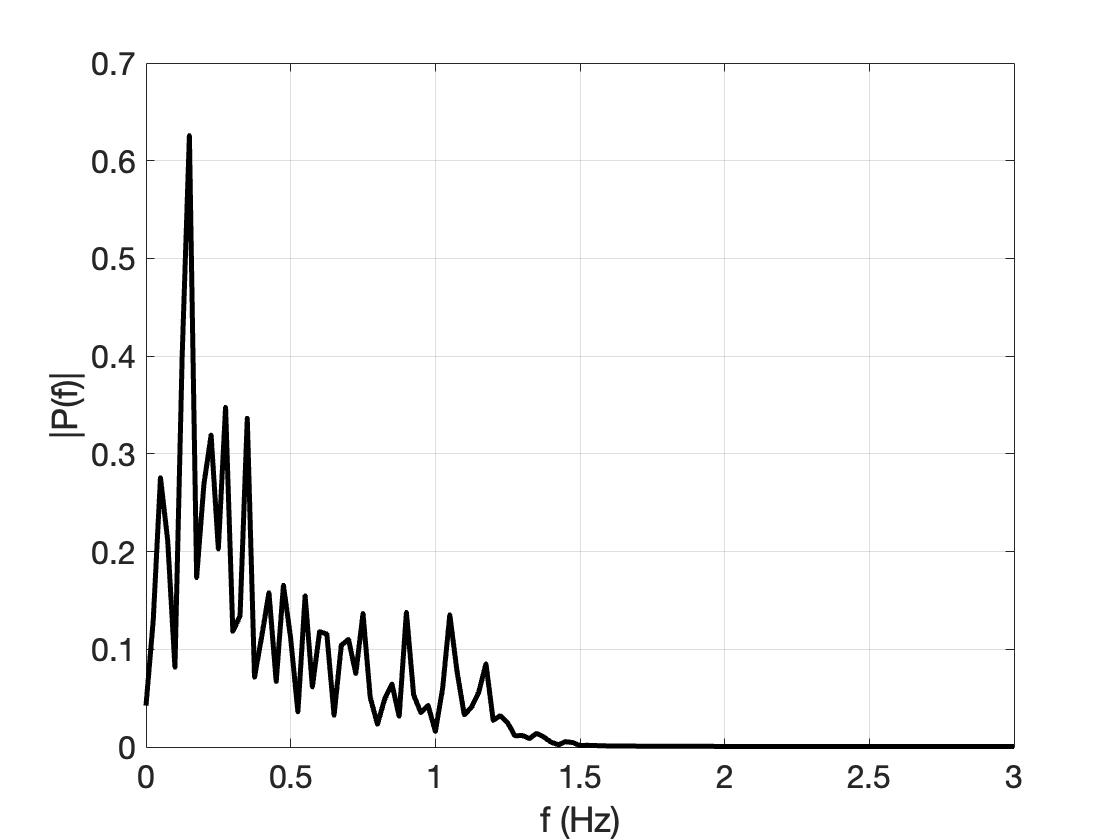}
\caption{Esquematic representation of the proposed algorithm.}
\label{fig::fftSpectrum}
\end{figure}
\begin{figure}
     \centering
     \begin{subfigure}[b]{0.3\textwidth}
         \centering
         \includegraphics[scale=0.4]{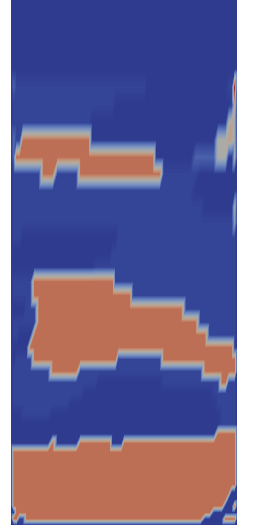}
         \caption{$v=0.14 \ m/s$}
         \label{fig:y equals x}
     \end{subfigure}
     \hfill
     \begin{subfigure}[b]{0.3\textwidth}
         \centering
         \includegraphics[scale=0.4]{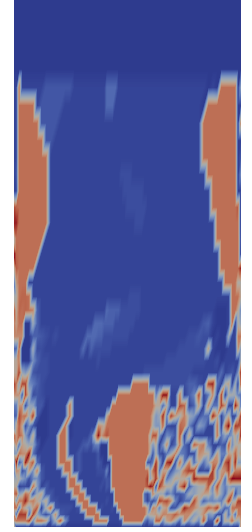}
         \caption{$v=0.50 \ m/s$}
         \label{fig:three sin x}
     \end{subfigure}
     \hfill
     \begin{subfigure}[b]{0.3\textwidth}
         \centering
         \includegraphics[scale=0.4]{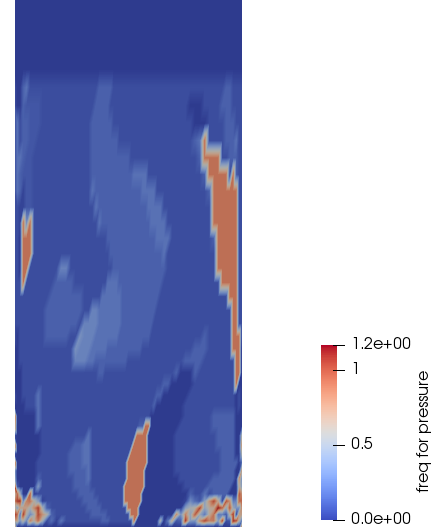}
         \caption{$v=0.90 \ m/s$}
         \label{fig:five over x}
     \end{subfigure}
        \caption{Mapping of the dominating frequency of the pressure field for different air inlet velocities using FFT.}
        \label{fig:threeGraphs}
\end{figure}

{
To illustrate the significance of selecting the correct analysis point, a mapping of all mesh points is conducted, and the signal is studied using FFT. Figure~\ref{fig:threeGraphs} shows the mapping result for the normalized pressure fields in three representative velocities, which are 0.14, 0.50, and 0.90 cm/s air inlet velocity. The frequency value corresponding to each location is determined by identifying the peak with the highest amplitude, as shown in the FFT figure. This analysis is performed for the normalized pressure field, which exhibits good behavior and is practically more accessible than other simulation variables.

An examination of the profiles reveals the variation in the distribution map of the predominant frequencies at each point within the liquid, particularly at the lowest velocity (a).
For higher velocities (b) and (c), more uniformity is evident, making it easier to select representative points for analysis.
Another critical aspect is that the reactor's characteristic reverse flow encourages the highest frequencies to be near the walls, which is visible only at high inlet velocities.
To detect this phenomenon, an analysis of the entire field was necessary, which would be unfeasible under experimental conditions.
This analysis helped us understand the distribution of the maximum frequencies in the system.
However, it was verified that the FFT analysis is primarily useful to evaluate local flow patterns and not global ones.
}

\subsection{HODMD analysis}
{
Table~\ref{tab::pressure_velocity_HODMD} represents the HODMD results for three values of the inlet velocities, showing the frequencies ($\omega$) in Hz, the damping rates ($\zeta$),
and amplitudes ($a$) relative to the expansion of each field as shown in Equation~\ref{b0}.
}

{
\color{blue}
\begin{table}[h!]
\centering
\caption{HODMD coefficients of the normalized pressure and velocity-$x$ component field data for different air inlet velocities.}
\begin{tabular}{|c|c|c|c|c|c|c|}
\hline
\multirow{ 2}{*}{$v$~[cm/s]} & \multicolumn{3}{|c|}{ \textbf{Normalized pressure}} & \multicolumn{3}{|c|}{ \textbf{Velocity-$x$}} \\
\cline{2-7}
           & $\omega_k$ & $-\zeta_k$ & $a_k$ & $\omega_k$ & $-\zeta_k$ & $a_k$ \\
\hline
& 0.0333 & 0.0111 & 0.1762 & 0.0348 & 0.0103 & 0.008\\
& 0.0737 & 0.0928 & 0.1393 & 0.9073 & 0.0113 & 0.0069\\
0.14 & 0.8955 & 0.0030 & 0.0855 & 0.0075 & 0.0027 & 0.0053\\
& 0.9024 & 0.0452 & 0.0638 & 0.0877 & 0.0145 & 0.0016\\
& 1.0077 & 0.0320 & 0.0322 & 0.8236 & 0.0226 & 0.001\\
& 0.8211 & 0.0217 & 0.0261 & - & - & - \\
\hline
& 0.0402 & 0.0044 & 0.2398 & 0.0382 & $9.78\times 10^{-5}$ & 0.0173\\
& 0.9007 & 0.0010 & 0.0921 & 0.1388 & 0.0541 & 0.0171 \\
0.50 & 0.8567 & 0.0449 & 0.0425 & 0.0699 & 0.0054 & 0.005 \\
& 0.9138 & 0.0166 & 0.0380 & 0.8941 & 0.0017 & 0.0048 \\
& 0.0957 & 0.0374 & 0.0374 & 0.178 & 0.0061 & 0.0042 \\
& - & - & - & 0.2124 & 0.0042 & 0.0029 \\
& - & - & - & 0.8467 & $9.19 \times 10^{-4}$ & 0.0022 \\
& - & - & - & 0.9298 & 0.0019 & 0.0013 \\
\hline
& 0.0632 & 0.0234 & 0.6101 & 0.0686 & 0.0037 & 0.0531 \\
0.90 & 0.2505 & 0.2584 & 0.3478 & 0.1207 & 0.0103 & 0.0092\\
& 0.0901 & 0.0086 & 0.2166 & 0.307 & 0.038 & 0.0087 \\
& 0.8890 & 0.0051 & 0.0470 & 0.9343 & 0.0049 & 0.0053 \\
& - & - & - & 0.8659 & 0.004 & 0.0044 \\
\hline
\end{tabular}
\label{tab::pressure_velocity_HODMD}
\end{table}
}

{
The HODMD, as mentioned, requires a calibration of the parameters; in particular, for the cases performed in this work, the value of the truncation $N$, $5\leq N\leq10$ and the value of $d$ between $5\leq d \leq20$,
which compared to other applications is significantly low \cite{BELTRAN}; this is an indication that in the studied time interval, the method can capture the spectral complexity
(number of modes that describes the system behavior) of the system. 
}

%The upper tables \ref{tab:Tab_SegementosD1} correspond to all the modes obtained by the algorithm proposed for the normalized pressure fields of the velocities $0.14, 0.50$, and $0.90 cm/s$. In contrast, the lower tables correspond to the same inflow velocities but velocity X field. These data are chosen to show how the results are obtained. 

The results show that the system is made up of several modes of different intensities, but the most relevant ones are those of lower frequency (between $0.03$ and $0.07~Hz$).
So also, the velocity and pressure fields show that the fundamental frequency is in the same order of magnitude and very close values (even to four decimal places of precision).
It is remarkable to show that the number of modes and frequencies are very similar for both velocity fields (for the same velocity).
This is a robustness indication, and above all, these modes represent a global response of the system, which is in concordance with the typically used approach in practice, that of using the FFT at specific points.
The fact that needing the entire field to obtain the important modes may be impractical for its application, but as seen in the following sections,
the proposed algorithm is applicable for a smaller number of data and provides robust results.

\subsection{DMD modes}
\begin{figure}
     \centering
     \begin{subfigure}[b]{0.3\textwidth}
         \centering
         \includegraphics[scale=0.4]{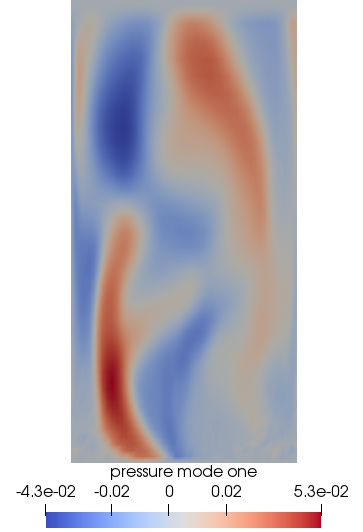}
         \caption{Pressure Mode one in the liquid region.}
          \label{fig:pmodeone}
     \end{subfigure}
     \hfill
     \begin{subfigure}[b]{0.3\textwidth}
         \centering
         \includegraphics[scale=0.4]{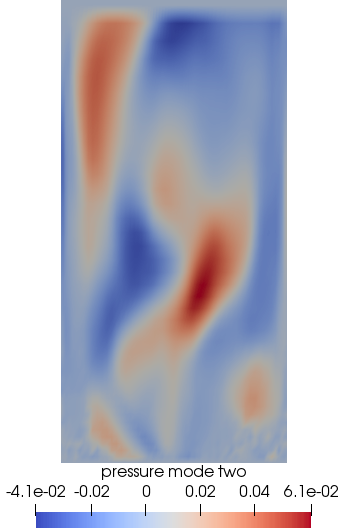}
         \caption{Pressure Mode two in the liquid region.}
          \label{fig:pmodetwo}
     \end{subfigure}
     \hfill
     \begin{subfigure}[b]{0.3\textwidth}
         \centering
         \includegraphics[scale=0.4]{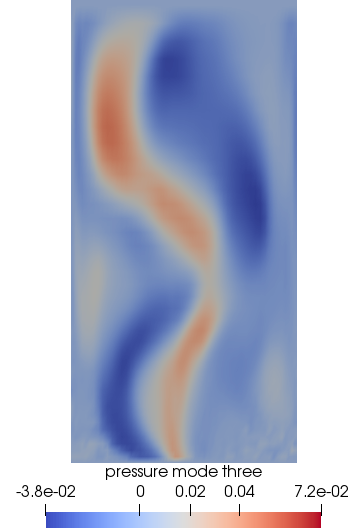}
         \caption{Pressure Mode three in the liquid region.}
          \label{fig:pmodethree}
     \end{subfigure}
        \caption{Pressure Mode for air inlet superficial velocity of $0.90$ $cm/s$ in the liquid region.}
        \label{fig:Pressuremode90}
\end{figure}

\begin{figure}
     \centering
     \begin{subfigure}[b]{0.3\textwidth}
         \centering
         \includegraphics[scale=0.4]{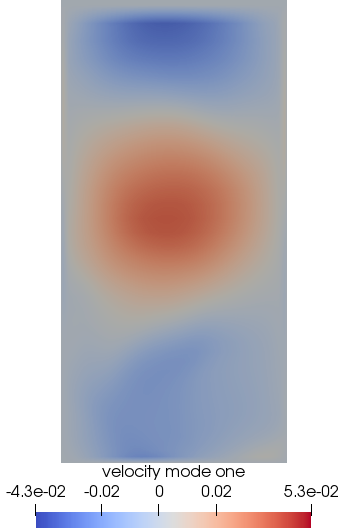}
         \caption{Velocity Mode one in the liquid region.}
          \label{fig:vmodeone}
     \end{subfigure}
     \hfill
     \begin{subfigure}[b]{0.3\textwidth}
         \centering
         \includegraphics[scale=0.4]{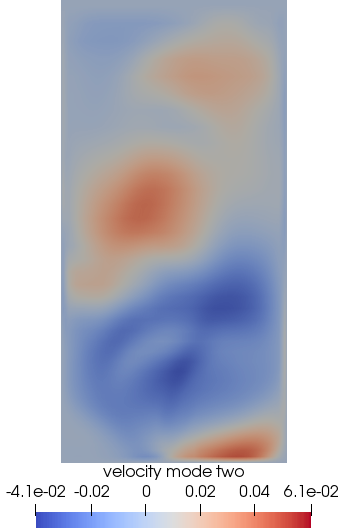}
         \caption{Velocity Mode two in the liquid region.}
          \label{fig:vmodetwo}
     \end{subfigure}
     \hfill
     \begin{subfigure}[b]{0.3\textwidth}
         \centering
         \includegraphics[scale=0.4]{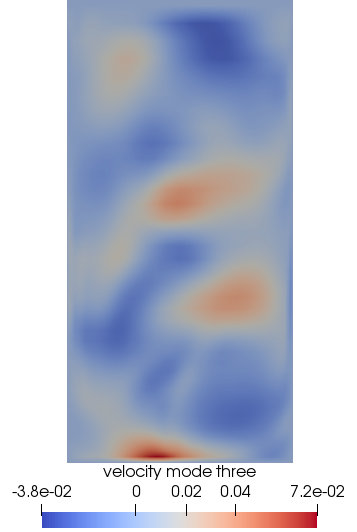}
         \caption{Velocity Mode three in the liquid region.}
          \label{fig:vmodethree}
     \end{subfigure}
        \caption{Velocity Mode for air inlet superficial velocity of $0.90$ $cm/s$ in the liquid region.}
        \label{fig:Velocitymode90}
\end{figure}

This section analyzes HODMD modes behavior for the pressure and velocity fields.
In Figures~\ref{fig:Pressuremode90} and~\ref{fig:Velocitymode90}, one can observe the three most crucial pressure and air $x$-velocity component modes for an inlet superficial velocity of $0.90~cm/s$.
Figure~\ref{fig:Pressuremode90} show that these three modes have similar magnitudes and ranges.
Although there are oscillations in the pressure field, which Figure~\ref{fig:isoPress} corroborates.
In the liquid region, the pressure has a slight variation (compared to the total pressure order of magnitude).
Likewise, the pressure change in the liquid domain, produced by momentum injection by the air, can be considered a harder parameter to identify the bubble column dynamics.
This result was expected since the pressure field is indeed governed by hydraulic pressure.

 Figure~\ref{fig:Velocitymode90} shows that the magnitude of the principal mode is similar but concentrated in three main regions (top, bottom, and center) of the liquid part. This represents the oscillation and dynamic effects of the air injection in the bubble column. In addition, one can see that the velocity magnitude is also in the order of magnitude of the superficial inlet velocity and the air domain velocity. Figure~\ref{fig:Velocitymode90} also indicates that only a few groups of points are needed to measure the fluid flow dynamics. One can select, for example, six points to collect velocity data (two in each region - center, top, and bottom) in order to capture the system fundamental frequency for different superficial air injection velocities.

\begin{figure}[!htb]
\centering
\includegraphics[scale=0.3] {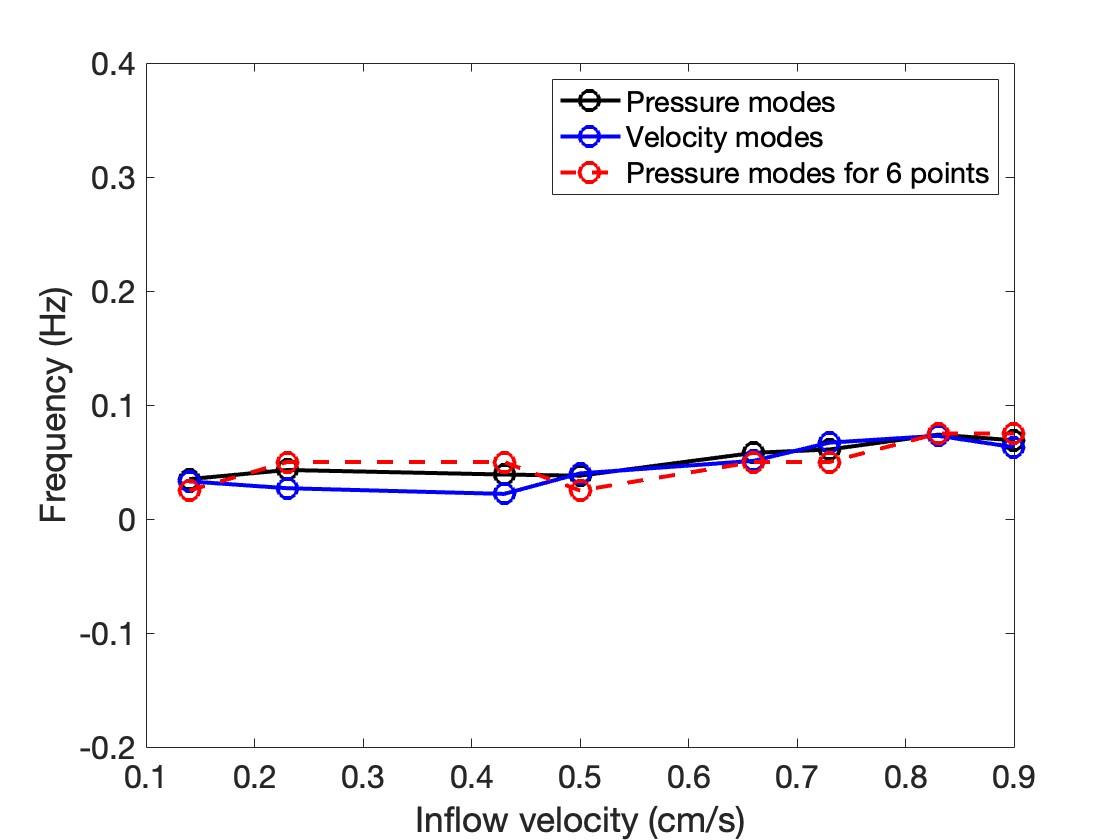}
\caption{ Dominant frequency over inflow velocity with six data points against the whole domain data collection}
\label{fig:fewpoints}
\end{figure}

Figure~\ref{fig:fewpoints} demonstrates a vital result related to the principal frequency of this system. The HODMD can capture the primary domain frequency with only six points for different inlet superficial air velocities. It means that without much acknowledgment of the flow field,  one can capture and/or reconstruct the whole bubble column dynamics, differently from what is usually obtained with the most commonly used method, FFT.   

%%%%%%%%%%%%%%%

%%%%%%%%5

\clearpage
\section{Conclusions} \label{sec:Conclusion}

{
In this manuscript, we present a comparative study of two approaches to predict the dynamics of a 2D bubble column system: the FFT and the HODMD.
It was demonstrated that traditional methods, such as FFT, may not always be effective in capturing the dynamic behavior of a complex dispersed multiphase flow system like a bubble column.
However, the HODMD method has proven to be a reliable and efficient data-driven approach for characterizing the system's dynamics and obtaining accurate results with a limited number of sampling points.

The data set comprised a comprehensive range of simulations of a 2D bubble column system, encompassing eight distinct operational conditions with a superficial velocity range between 0.14 and 0.9 $cm/s$.
From these simulations, we observed, by an FFT analysis, that depending on the points where we apply the FFT, the result would miscalculate the dominant frequency of the system.
On the other hand, employing the HODMD, we capture the system's accurate dominant frequency and modes.

These results show that HODMD is adequate to capture the dynamic of the system.
Besides, we used six arbitrarily selected points over the bubble column instead of the whole domain points and then applied HODMD.
Our results showed that FFT could not adequately describe the system as it has been done for a long time in the industry.
However, with a few sampling points, HODMD can well represent and reconstruct the dynamics of this complex dispersed multiphase flow system.
It is a significant result because one could rebuild the dynamic of the experimental bubble column with a few points without the numerical simulations.
}

%The results of this study demonstrate the importance of data-driven methods in understanding the dynamics of multiphase flow systems, and could contribute to the development of more effective models for optimizing the performance of bubble columns.
%Hence, our work is relevant to researchers and practitioners working in the field of multiphase flow dynamics and could stimulate future investigations in this area.

\section*{Acknowledgements}
This work was supported by the National Oil Agency, ANP-Brazil (2015/00350-7). 

\clearpage

\bibliography{paperbib}

\end{document}